\documentclass[prl,showpacs,floatfix,twocolumn]{revtex4}

\usepackage{graphicx}
\usepackage{dcolumn}
\usepackage{float}
\usepackage{amsmath}

\newcommand{\comment}[1]{}

\begin{document}

\title{\boldmath  Ultrapure Multilayer Graphene in Bromine Intercalated Graphite \unboldmath}

\author{J. Hwang$^{1}$}\email{jhwang@pusan.ac.kr} \author{J. P. Carbotte$^2$} \author{S. Tongay$^3$}\altaffiliation[Current address: ]{Department of Materials Science and Engineering, University of Florida, Gainesville, Florida 32611, USA.}
\author{A. F. Hebard$^3$}  \author{D. B. Tanner$^3$}

\affiliation{$^{1}$Department of Physics, Pusan National University, Busan 609-735, Republic of Korea\\
$^{2}$Department of Physics and Astronomy, McMaster University, Hamilton, ONL8S 4M1, Canada\\
$^{3}$Department of Physics, University of Florida, Gainesville, FL 32611, USA}

\date{\today}


%
%
\begin{abstract}
We investigate the optical properties of bromine intercalated highly orientated pyrolytic graphite (Br-HOPG) and provide a novel interpretation of the data. We observe new absorption features below 620 meV which are absent in the absorption spectrum of graphite. Comparing our results with those of theoretical studies on graphite, single and bilayer graphene as well
as recent optical studies of multilayer graphene, we conclude that Br-HOPG contains the signatures of ultrapure bilayer, single layer graphene, and graphite. The observed supermetallic conductivity of Br-HOPG is identified with the presence of very high mobility ($\simeq$ 121,000 cm$^2$V$^{-1}$s$^{-1}$ at room temperature and at very high carrier density) multilayer graphene components in our sample. This could provide a new avenue for single and multilayer graphene research.
\end{abstract}


\maketitle
%

Recently several allotropes of carbon including Buckminster-fullerenes, carbon nanotubes, and graphene have been discovered and have attracted much attention\cite{kroto85,iijima91,novoselov04,zhang05,geim07,wang09,li08}. These carbon-based materials have unique crystal structures and interesting and useful physical, chemical, and mechanical properties, leading to potential applications in various fields including carbon based electronics. Multilayer graphene has become especially important because of the known unique Dirac cone band structure in single layer graphene with its chirality, non-trivial Berry phase, half integral Hall plateaus etc \cite{novoselov04,zhang05,geim07,wang09,li08}. The parent material for graphene is graphite, which has a layered structure. Each layer of graphite consists only of carbon with hexagonal lattice structure. The in-plane carbon-carbon interaction provides a strong covalent bond and inter-plane interaction is a weak van der Waals force. One can easily insert atoms or molecules between the sheets in graphite through intercalation processes. In most cases Li, Ca, and K have been used for intercalation\cite{dresselhaus81,boehm94,emery07,chakraborty07}. Studies of Br intercalated graphite mostly focused on the structural properties of the intercalant layer itself\cite{kortan82,erbil83}. For our considerations here it is important to note that the c-axis conductivity at high Br concentrations is reduced by more than factor of 5 over its value in graphite\cite{dresselhaus81,tongay10} while for K it is increased by a factor of $\sim$ 25. Thus Br$_2$ leads to more electrically isolated graphite planes (two dimensional graphene) while K integrates them into a more three dimensional electronic structure.

In this study, we intercalated bromine molecules between graphite sheets by exposing highly oriented pyrolytic graphite (HOPG) to a bromine gas environment. We prepared three bromine intercalated samples with different exposed time, 15 minutes, 30 minutes, and 100 minutes, respectively, labeled as 15-min Br-HOPG, 30-min Br-HOPG, and 100-min Br-HOPG, respectively. In general, the intercalation can introduce charge carriers in the graphite sheets and can make the distance between sheets larger. However, it is not clear how uniformly Br atoms are distributed between the graphite layers and how much charge transfer occurs. It is known that Br-HOPG shows carrier density increases, supermetallic behavior, and reduced scattering rate at high intercalation levels\cite{tongay10,thompson76}. The reduction of the scattering rate at high Br intercalation is yet to be understood. Here, we focus on this issue and on additional new features, which appear below 620 meV in the optical spectra of Br-HOPG.

\begin{figure}[t]
  \vspace*{-1.0 cm}%
  \centerline{\includegraphics[width=3.0 in]{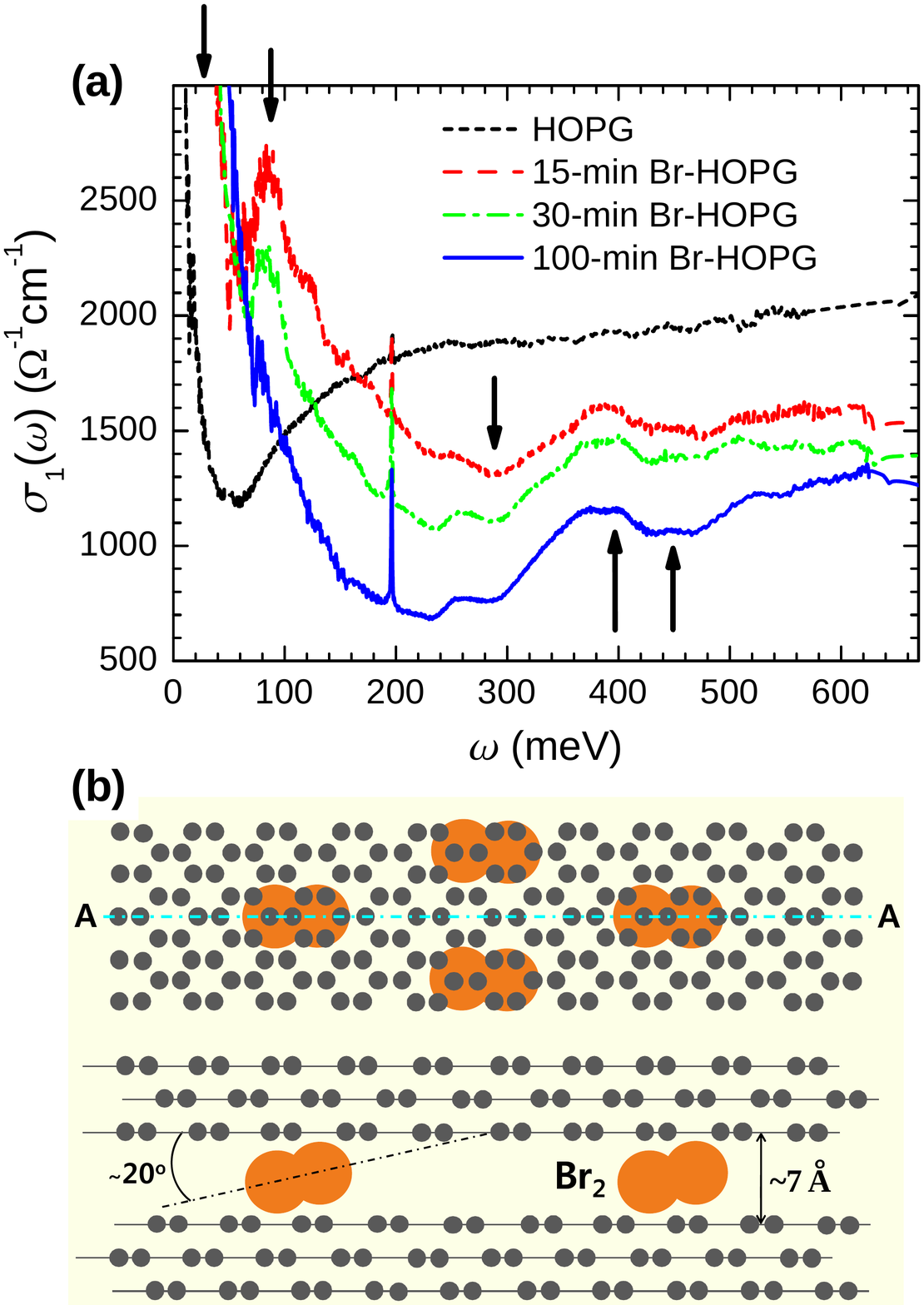}}%
  \vspace*{-0.5 cm}%
\caption{(Color online) (a) The room temperature conductivity spectra of bare HOPG and three Br-HOPG with different intercalation times. We indicate the five differences between bare HOPG and Br-HOPG with vertical arrows (see in the text). (b) We show a schematic diagram of Br$_2$ intercalated graphite: (upper) a top (in-pane) view and (lower) a side view (along c-axis) of A-A line cross section. This diagram is based on results by Erbil {\it et al.}\cite{erbil83}.}
  \label{fig1}
\end{figure}

In Fig. 1 (a) we show the in-plane optical conductivities of four samples including bare HOPG, obtained from the measured reflectance via Kramers-Kronig analysis~\cite{wooten72}. The optical conductivity of the HOPG at room temperature agrees well with that previously reported\cite{li06,kuzmenko08}. The optical conductivities of the Br-HOPG samples show dramatic changes as compared with HOPG. Some of the properties of Br-HOPG have already been reported, such as the supermetallicity, arising from a high Drude plasma frequency and a scattering rate reduction\cite{tongay10}. The spectra show the appearance of new absorption features in mid-infrared region (marked with arrows in the figure) and an overall suppression of the conductivity in the spectral region from mid-infrared through near infrared with increasing Br intercalation.

To understand the conductivity suppression of Br-HOPG in mid-infrared and near infrared region we take the optical conductivity of bare HOPG near 620 meV (the relatively flat part of the spectrum) as a reference. This reference is the parallel contribution to the conductivity of each graphite layer separately\cite{kuzmenko08}. We normalize the conductivity at 620 meV to $G_{s0}$, the effective optical conductance of a single graphite sheet (graphene) in HOPG. The normalized conductance can be written as $G(\omega)/G_{s0} = d_{c}\:\sigma_1(\omega)$, where $d_c$ is the c-axis lattice constant and $\sigma_1(\omega)$ is the real part of the optical conductivity. In bare HOPG $d_c$ = 3.4 \AA. In the Br-treated samples, the normalized conductance yields the c-axis spacing, because it is related to the number density of graphite sheets along the stacking axis. In the inset of Fig. 2 (a) we display the extracted c-axis constant, $d_c$ as a function of (Br-intercalation time)$^{1/2}$ for all four samples; values are scaled by the c-axis lattice constant of bare HOPG. The c-axis lattice constant increases as the amount of Br intercalants increases, because the Br atoms go between the graphite layers and push the graphite sheets further apart as sketched in Fig. 1 (b). In the early stage of intercalation (up to 30 minutes), $d_c$ increases as the square root of time, as found in Ref.~\cite{tongay10} for Br uptake, and then begins to saturate reaching $d_{c} \simeq 5.2$ \AA\ for the 100-min sample. This spacing is smaller than the 7.0 \AA\ Br-layer spacing found in structural studies\cite{erbil83} but bigger than the average of 4.4 \AA. Our effective c-axis lattice constant is an average spacing. We can not tell whether the Br atoms go between every graphite sheet or not.

It is interesting to compare further the conductance of our Br intercalated samples with the conductance of HOPG shown in Fig. 1 (a). We emphasize five differences: a broader Drude mode, a peak near 85 meV, a reduction of spectral weight near 300 meV, a new peak near 400 meV, and a valley near 450 meV. The peak around 85 meV in the 15-min sample (red dashed line) shifts to lower energies and appears to vanish in the 100-min sample (blue solid line). The reduction at higher photon energies below roughly 300 meV  is accompanied by increased optical spectral weight in the Drude spectrum. The new peak near $\omega =$ 400 meV is common to all three doped samples. This peak is followed by the valley before the conductance rises to its constant uniform background at high energies. To gain insight into the possible origin of these five features it is useful to recall recent theoretical and experimental studies of the conductivity of single\cite{gusynin06,gusynin07,li08,kuzmenko08} and double layered graphene\cite{nicol08,zhang08,li09,kuzmenko09,kuzmenko09a}.

For a single layer in the clean limit a remarkably simple formula of the conductance, $G_s(\omega)$, was obtained \cite{gusynin06,gusynin07,gusynin09} namely $\frac{G_{s}(\omega)}{G_{s0}}=4|\mu|\:\delta(\omega)+\Theta\Big{(}\frac{\omega}{2}-|\mu|\Big{)}$
where $G_{s0} \equiv \pi e^2/(2 h)$ is the conductivity of a single graphene sheet, $e$ is the charge on the electron, and $h$ is the Plank's constant. When broadening is included, the Dirac delta function in the first term is replaced by a Drude spectrum. The second term is a Heaviside function giving a constant universal background for photon energy $\omega \geq 2|\mu|$ with $\mu$, the chemical potential describing charging. The optical spectral weight under the Drude spectrum which comes from the intraband transitions is exactly equal to the missing weight under the background, which starts at $2|\mu|$ and has its origin in the interband transitions from lower to upper Dirac cones. Identification of this onset allows a measure of $|\mu|$ and hence of the charge transfer from the Br to the graphite planes.

For the bilayer a slightly more complicated formula for the conductance, $G_{b}(\omega)$, can be derived\cite{nicol08} and is
\begin{eqnarray}
\frac{G_{b}(\omega)}{G_{b0}}&=&\Big{[}\frac{\omega+2\gamma}{2(\omega+\gamma)}+\frac{\omega-2\gamma}
{2(\omega-\gamma)}\Big{]}\Theta(\omega-2|\mu|)\nonumber \\
&+&\!\!\frac{\gamma^2}{2\omega^2}\Big{[}\Theta(\omega\!-\!2|\mu|\!-\!\gamma)
+\Theta(\omega\!-\!2|\mu|\!+\!\gamma)\Big{]}\Theta(\omega\!-\!\gamma)\nonumber \\
&+&A(\mu)\delta(\omega)+B(\mu)\delta(\omega-\gamma),
\end{eqnarray}
with $ A(\mu) \equiv \frac{4|\mu|(|\mu|+\gamma)}{2|\mu|+\gamma}+\frac{4|\mu|(|\mu|-\gamma)}{2|\mu|-\gamma}
\Theta(|\mu|-\gamma)$ and $B(\mu) \equiv \frac{\gamma}{2}\Big{[}\ln\frac{2|\mu|+\gamma}{\gamma}-\ln\frac{2|\mu|-\gamma}{\gamma}\Theta(|\mu|-\gamma)\Big{]}$,
where $G_{b0} = 2G_{s0} = e^2/2\hbar$, which is twice the conductance of a single graphene sheet. The new feature is that now the hopping parameter between planes, $\gamma$, enters. The absorption peak (a Delta function in the clean limit) at $\omega = \gamma$ is a result of the energy splitting between the bands. One set of bands is displaced with respect to the other by $\gamma$ and at finite $\mu$ this leads to a {\it degeneracy} of optical transitions between the split bands. For graphite\cite{johnson73,dillon77,partoens06} $\gamma =$ 400 meV and we identify this new absorption with a peak in our data for $\sigma_1(\omega)$ seen in all three Br intercalated samples around 400 meV. As the Br is intercalated the system acquires some of the characteristics associated with bilayer graphene.

In Fig. 3 (a) we show numerical results, $[G_{s}(\omega)/G_{s0}+G_{b}(\omega)/G_{b0}]/2$, based on equations (1) and (2) for an equally weighted sum of a single and a doubled layer of graphene for several values of the chemical potential. In all cases there is a Drude spectrum at small $\omega$, a constant universal conductance background at large $\omega$ and a sharp rise in absorption at $\omega = 2|\mu|$. There is also a peak at $\omega=\gamma$, the energy shift between the two sets of bands in the bilayer. When twice the chemical potential $2|\mu|$ is greater than $\gamma$ there is also a region of depressed conductivity between $\gamma$ and $2|\mu|$. All these features are seen in the blue solid curve of Fig. 2 (a) for the most Br doped case and we identify the chemical potential as $\sim$ 260 meV = 0.65$\gamma$.

\begin{figure}[t]
  \vspace*{-1.2 cm}%
  \centerline{\includegraphics[width=3.0 in]{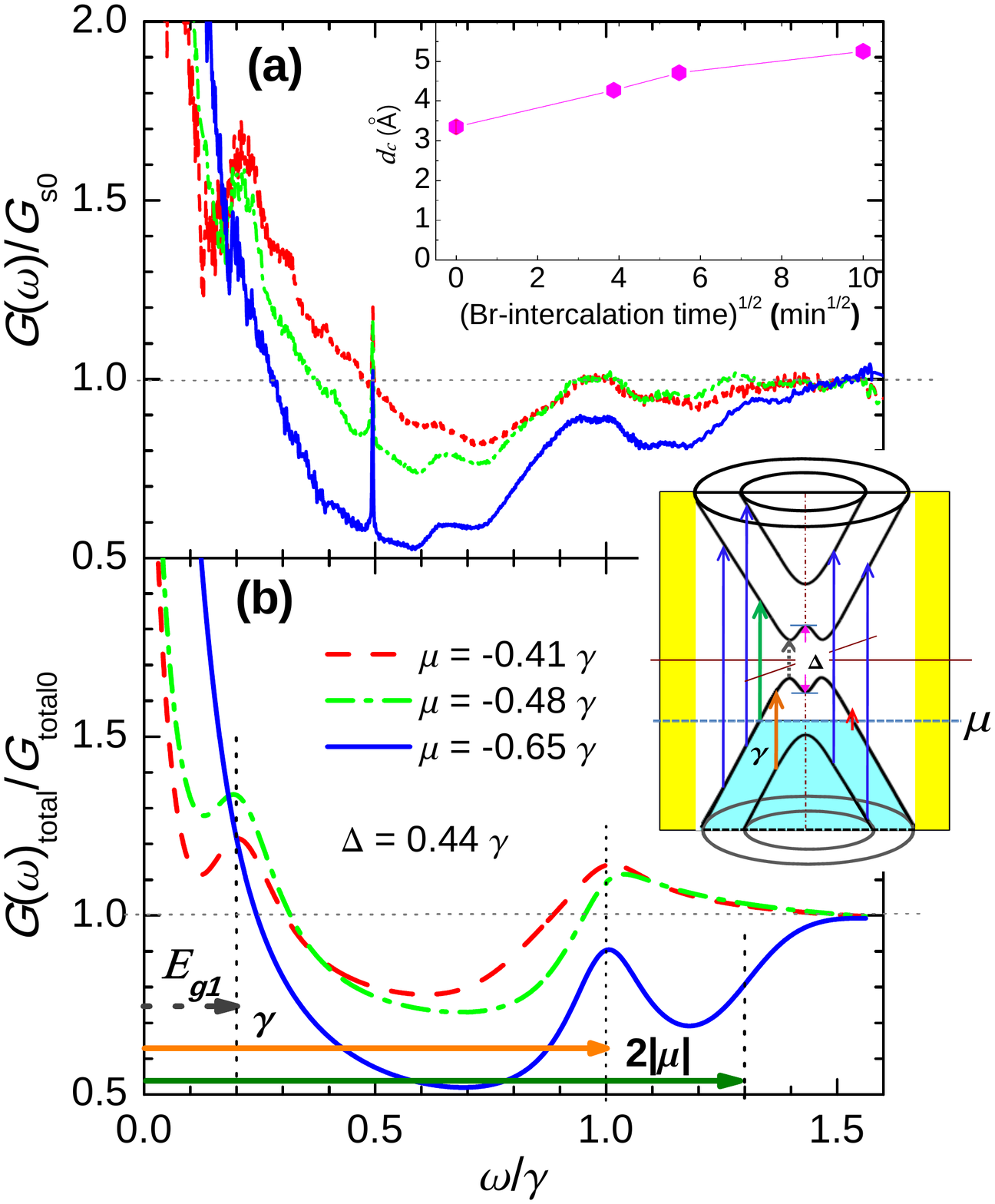}}%
  \vspace*{-1.5 cm}%
\caption{(Color online) (a) Room temperature optical conductivity normalized at 620 meV, as described in Ref.~\cite{kuzmenko08}. This normalization allow the extraction of an effective c-axis lattice constant, shown in the inset (see in the text). (b) Simulation of our data using our model calculation. (See the text for a detailed description.) In the inset we show a schematic diagram of a band structure of bilayer graphene with an anisotropy gap, $\Delta \cong 0.44 \gamma$ (not to scale). Here, the gap is caused by a built-in potential difference between upper and lower layers. The red arrow shows the intraband Drude conductivity, the blue the interband transitions, the orange the $\gamma$ transition, and the dark green the onset of the interband transition. The gray dotted arrow shows the transition between the minimum gap, $E_{g1}$. We note that this transition would not appear at low temperature or in a highly doped system.}
 \label{fig2}
\end{figure}

A qualitative understanding of the data for two other samples with less Br intercalated requires further elaboration of the results for bilayer graphene. When the two layers in the bilayer are not equivalent as would be expected if the Br intercalation process is not the same on the two surfaces of the bilayer, there can be an anisotropy gap, $\Delta$. Details of the theory can be found in reference \cite{nicol08} but here it is sufficient that we model the main results. When a gap develops between bands that would otherwise be degenerate, part of the Drude spectrum will become gapped. We use this idea in the results of our own simulations shown in Fig. 2 (b). We add the conductance of three subcomponents: graphite taken as measured in our experiment and an equal mixture of single and bilayer graphene. $G_{total}(\omega)/G_{total0} \equiv f_{g}G_{g}(\omega)/G_{g0}+(1-f_{g})[G_{s}(\omega)/G_{s0}+G_{b}(\omega)/G_{b0}]/2$, where $G_{total0}$ is $G_{s0}\times\mbox{the number of total layers}$, $f_{g}$ is the graphite fraction, and $G_{g}/G_{g0}$ is the normalized conductance of the bare HOPG. We used $f_{g}$ = 0.70, 0.50, and 0.38 for Br 15-min, 30-min, and 100-min samples, respectively. These graphite fractions are consistent with the values estimated independently using the effective c-axis constant $d_{c} \cong f_g \:d_{c0} + (1-f_g) \:d_{c-sat}$ (see Fig. 2 (a)), where $d_{c0}$ is the c-axis constant (3.4 \AA) of graphite and $d_{c-sat}$ is the graphite-bromine-graphite sandwich width ($\simeq 7.0$ \AA)\cite{erbil83}: $f_g = $0.75, 0.63, and 0.48, respectively. We also shifted the chemical potential for first and second curves to account for an overall reduction in Br, but kept the same ratio between ionized and non-ionized as found for the most Br rich sample so the $\mu$'s were set to 0.41$\gamma$, 0.48$\gamma$, and 0.65$\gamma$, respectively. Finally for the bilayer component in each case we included an anisotropy gap of $\Delta = 0.44 \gamma$. To do this, in equation (2) we replace the Drude term $A(\mu)\delta(\omega)$ with a two component term $f_{gap}A(\mu)\delta(\omega-E_{g1})+(1-f_{gap})A(\mu)\delta(\omega)$ and $f_{gap}$ = 0.8, 0.5, and 0.0, for Br 15-min, 30-min, and 100-min samples, respectively, where $f_{gap}$ in the portion of gaped states and $E_{g1}(\equiv \gamma\Delta /(2\sqrt{\Delta^2+\gamma^2}))$~\cite{nicol08} is the minimal gap between upper and lower bands (see Fig. 2 (b)). These percentages roughly represent the idea that as doping proceeds the mechanism for anisotropy gap formation because less effective as the top and bottom surface of a given bilayer should experience more and more the presence of the same Br coating.

\begin{figure}[t]
  \vspace*{-1.0 cm}%
  \centerline{\includegraphics[width=3.0 in]{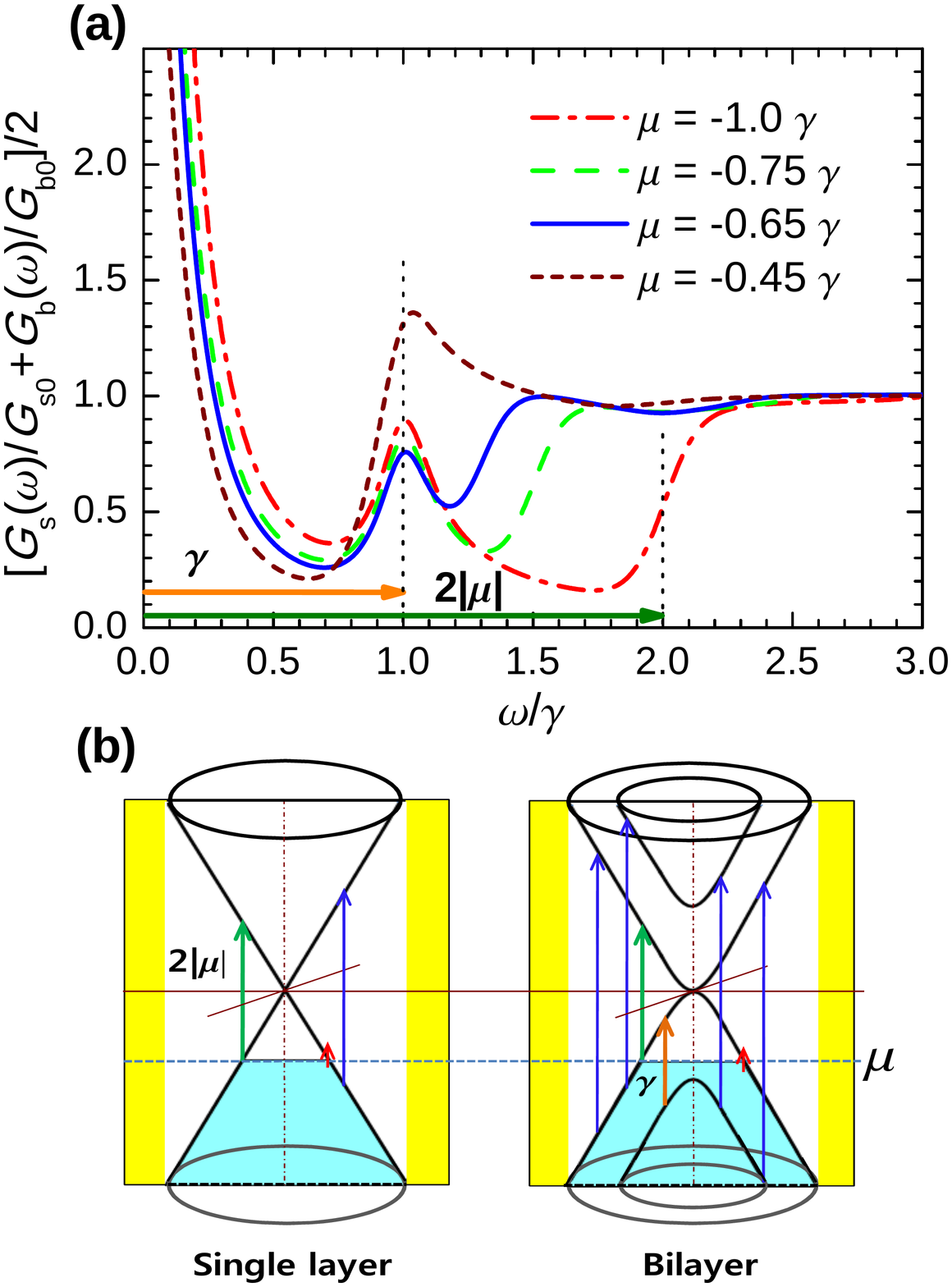}}%
  \vspace*{-0.9 cm}%
\caption{(Color online) (a) Numerical calculations based on Eq. (1) and (2) for an equally weighted sum of single and bilayer graphene for several chemical potentials. We include the temperature broadening in the calculation. The orange horizontal arrow shows the degenerate $\gamma$-transition. The dark green arrow the onset of the interband transition for $\mu = -1.0 \gamma$. (b) We show schematic diagrams of band structures of both single and bilayer graphene. The red arrow shows the intraband Drude transition, the blue the interband transitions, the dark green the onset of these transitions, and the orange the degenerate $\gamma$-transition.}
 \label{fig3}
\end{figure}

The information on the chemical potentials allows us to estimate the charge transfer to the graphite sheets. Via weight uptake measurements, we know the amount of Br intercalated versus the time of exposure to the Br atmosphere: 2.3, 3.3, and 6.0 at \% Br for 15-min, 30-min, and 100-min samples, respectively\cite{tongay10}. The bandwidth is $W \simeq$ 7.0 eV. Because we have a linear band structure (Dirac cone) near the Fermi Surface, the number of filled states is roughly proportional to $|\mu|^2$, where $\mu$ is the chemical potential. For our 100-min sample, if all intercalated Br are ionized the chemical potential would be $|\mu_{assumed}| = \sqrt{0.06}\:W \cong 1.72$ eV. The extracted real chemical potential from the fit is $|\mu_{real}| = 0.65 \gamma \cong 0.26$ eV, which comes from the ionization. The ionization rate is $(|\mu_{real}|/|\mu_{assumed}|)^2 = 0.023$ or 2.3 \%. We also calculate the ionization rates for other two lower doped samples (15-min and 30-min) and get 2.4 \% and 2.2 \%, respectively. The ionization rate can be determined in a different way by using the measured change in plasma energy\cite{tongay10} and the effective c-axis lattice constant, $d_c$. The density of bare HOPG is roughly $\rho_{m}$ = 2.16 g/cm$^{3}$. By intercalation the density of carbon will be reduced to $\rho_m \times d_{c0}/d_c$. For our 100-min sample, the charge carrier density is $N_{Br^{-}}\cong 10^{20}$ holes/cm$^{3}$\cite{tongay10} and the effective c-axis constant is $d_c \cong 5.2 $~\AA. The density of carbon is 2.16 g/cm$^{3}\times3.4$~\AA/5.2 \AA${\relax} = 1.41$ g/cm$^{3}$, and the corresponding number of carbon atoms per cm$^{3}$ is 1.41 g/12.0 g $\times$ $N_{A}$ = 7.08$\times 10^{22}$ cm$^{-3}$, where $N_A$ is the Avogadro's number. The total number of intercalated Br atoms is $N_{Br} = 0.06\times(7.08\times 10^{22}) = 4.25\times 10^{21}$ cm$^{-3}$. The ionization rate is $N_{Br^{-}}/N_{Br} = 0.024$ or 2.4 \%. For other two samples (15-min and 30-min) the estimated ionization rates are 3.0 \% and 2.6 \%, respectively. The two approaches give very similar results, and agree with earlier results.\cite{platts77}

We also estimate the conductance of the single layer graphene in our samples using the fitting parameters, the measured dc resistivities, and the measured carrier density\cite{tongay10}: $G_{sample} = f_g G_g + (1-f_g) [G_s + G_b]/3 \cong f_g G_g + (1-f_g) G_s$. So $G_s = [G_{sample}-f_g G_g]/(1-f_g)$. For 100-min sample at room temperature since $G_{sample} = 6.3\times10^{-2}$ $ [\Omega^{-1}]$, $G_g = 0.1\times10^{-2}$ $[\Omega^{-1}]$, and $f_g = 0.38$, the conductance of the single layer graphene, $G_s = 1.01\times10^{-1}$ $[\Omega^{-1}]$. The corresponding mobility, $G_s/(n_{2d}\: e)$, is 121,000 cm$^{2}$V$^{-1}$s$^{-1}$, where $n_{2d}$ is the two dimensional carrier density, 5.2$\times$10$^{12}$ cm$^{-2}$ for the 100-min sample. The estimated mobility is about 10 times larger than that of single layer graphene fabricated on a substrate\cite{novoselov07} and is comparable to or slightly better than that of suspended single layer graphene\cite{bolotin08} even though it is at higher temperature (300 K) and at much higher carrier density. Interestingly, the graphene in our 100-min Br-HOPG sample shows higher carrier mobility than any reported values; it is "{\it ultrapure}" graphene.

In conclusion we investigated bromine-intercalated highly-oriented pyrolytic graphite (Br-HOPG) using infrared and optical spectroscopy. New features in the optical spectra of Br-HOPG, absent in the optical spectrum of HOPG, are observed. The characteristic energies of these new features, when compared with published optical data for graphene, graphite, and especially bilayer graphene, allow us to identify them as characteristic of the bilayer. Additionally, the chemical potential obtained from the simulation tells us that only a few percent of the intercalated Br molecules are ionized. The mobility in the isolated multilayer graphene components is found be very high and provides an explanation for the observed supermetallicity of our samples. This is a particularly attractive possibility considering the present excitement in the field of graphene and related materials\cite{novoselov04,zhang05,geim07,wang09,li08}. Measurements of the optical conductivity of highly intercalated samples under a magnetic field perpendicular to the layers would be particularly interesting.
We believe that our results may open a new avenue to the study of the properties of multilayered graphene.
%
%
\acknowledgments This work was supported by the National Research Foundation of Korea through grant No. 20100008552 (JH), by the NSERC (Canada) and the Canadian Institute for Advanced Research (CIFAR) (JC), by the NSF through grant DMR-1005301 (ST and AFH), and
by the DOE through grant DE-FG02-02ER45984 (DBT).

%
%

\end{document}